\documentclass[draft]{agujournal2019}
\usepackage{url} %this package should fix any errors with URLs in refs.
\usepackage{lineno}
\usepackage[finalnew]{trackchanges} %for better track changes. finalnew option will compile document with changes incorporated.
\usepackage{soul}
\usepackage{multirow}
\usepackage{rotating}
\usepackage{adjustbox}
\usepackage{graphicx}

\draftfalse

\journalname{JGR: Planets}

\begin{document}

%% ------------------------------------------------------------------------ %%
%  Title
%
% (A title should be specific, informative, and brief. Use
% abbreviations only if they are defined in the abstract. Titles that
% start with general keywords then specific terms are optimized in
% searches)
%
%% ------------------------------------------------------------------------ %%

% Example: \title{This is a test title}

\title{Constraining low-altitude lunar dust using the LADEE-UVS data}

%% ------------------------------------------------------------------------ %%
%
%  AUTHORS AND AFFILIATIONS
%
%% ------------------------------------------------------------------------ %%

% Authors are individuals who have significantly contributed to the
% research and preparation of the article. Group authors are allowed, if
% each author in the group is separately identified in an appendix.)

% List authors by first name or initial followed by last name and
% separated by commas. Use \affil{} to number affiliations, and
% \thanks{} for author notes.
% Additional author notes should be indicated with \thanks{} (for
% example, for current addresses).

% Example: \authors{A. B. Author\affil{1}\thanks{Current address, Antartica}, B. C. Author\affil{2,3}, and D. E.
% Author\affil{3,4}\thanks{Also funded by Monsanto.}}

\authors{H. Sharma\affil{1}, M. M. Hedman\affil{1}, and D. H. Wooden\affil{2}, A. Colaprete\affil{2}, A. M. Cook\affil{2}}

% \affiliation{1}{First Affiliation}
% \affiliation{2}{Second Affiliation}
% \affiliation{3}{Third Affiliation}
% \affiliation{4}{Fourth Affiliation}

\affiliation{1}{University of Idaho, Moscow ID, USA}
\affiliation{2}{NASA Ames Research Center, Moffett Field CA, USA}
%(repeat as many times as is necessary)

%% Corresponding Author:
% Corresponding author mailing address and e-mail address:

% (include name and email addresses of the corresponding author.  More
% than one corresponding author is allowed in this LaTeX file and for
% publication; but only one corresponding author is allowed in our
% editorial system.)

% Example: \correspondingauthor{First and Last Name}{email@address.edu}

\correspondingauthor{Himanshi Sharma}{shar9969@vandal.uidaho.edu}

%% Keypoints, final entry on title page.

%  List up to three key points (at least one is required)
%  Key Points summarize the main points and conclusions of the article
%  Each must be 140 characters or fewer with no special characters or punctuation and must be complete sentences

% Example:
% \begin{keypoints}
% \item	List up to three key points (at least one is required)
% \item	Key Points summarize the main points and conclusions of the article
% \item	Each must be 140 characters or fewer with no special characters or punctuation and must be complete sentences
% \end{keypoints}

\begin{keypoints}
\item LADEE'S UV Spectrometer made a series of observations that enable
us to probe the Moon's dust atmosphere at altitudes between 1 and 10 km
\item These observations are filtered to isolate dust signals and are compared to predicted dust atmospheres 
\item For a dust population with power law size distribution of index -3 and scale height 1 km, the upper limit on the dust density is 142 m$^{-3}$.
\end{keypoints}

%% ------------------------------------------------------------------------ %%
%
%  ABSTRACT and PLAIN LANGUAGE SUMMARY
%
% A good Abstract will begin with a short description of the problem
% being addressed, briefly describe the new data or analyses, then
% briefly states the main conclusion(s) and how they are supported and
% uncertainties.

% The Plain Language Summary should be written for a broad audience,
% including journalists and the science-interested public, that will not have 
% a background in your field.
%
% A Plain Language Summary is required in GRL, JGR: Planets, JGR: Biogeosciences,
% JGR: Oceans, G-Cubed, Reviews of Geophysics, and JAMES.
% see http://sharingscience.agu.org/creating-plain-language-summary/)
%
%% ------------------------------------------------------------------------ %%

%% \begin{abstract} starts the second page

\begin{abstract}
Studying lunar dust is vital to the exploration of the Moon and other airless planetary bodies. The Ultraviolet and Visible Spectrometer (UVS) on board the \add{Lunar Atmosphere and Dust Environment Explorer} (LADEE) spacecraft conducted a series of Almost Limb activities to look for dust near the dawn terminator region. During these activities the instrument stared at a fixed point in the zodiacal background off the Moon's limb while the spacecraft moved in retrograde orbit from the sunlit to the unlit side of the Moon. The spectra obtained from these activities probe altitudes within a few kilometers of the Moon's surface, a region whose dust populations were not well constrained by previous remote-sensing observations from orbiting spacecraft. Filtering these spectra to remove a varying instrumental signal enables constraints to be placed on potential signals from a dust atmosphere. These filtered spectra are compared with those predicted for dust atmospheres with various exponential scale heights and particle size distributions to yield upper limits on the dust number density for these potential populations. For a differential size distribution proportional to $s^{-3}$ (where $s$ is the particle size) and a scale height of 1 km, we obtain an upper limit on the number density of dust particles at the Moon's surface of 142 $m^{-3}$.
\end{abstract}

\section*{Plain Language Summary}
The Moon has a tenuous atmosphere of dust particles. It is important to study this dust environment in order to develop solutions for dust-related problems with exploration of the lunar surface, as well as understand the effects of dust transport on other airless planetary bodies. Lunar Atmosphere and Dust Environment Explorer's Ultraviolet and Visible Spectrometer made a series of Almost Limb observations that enable us to probe the Moon's dust atmosphere at altitudes ranging from 1 and 10 km above the surface. Data from these observations are processed and compared to the predicted signals from a variety of different dust populations. These comparisons yield new upper limits on dust density within a few kilometers of the lunar surface, providing additional constraints on dust population at low altitudes.

%% ------------------------------------------------------------------------ %%
%
%  TEXT
%
%% ------------------------------------------------------------------------ %%

%%% Suggested section heads:
% \section{Introduction}
%
% The main text should start with an introduction. Except for short
% manuscripts (such as comments and replies), the text should be divided
% into sections, each with its own heading.

% Headings should be sentence fragments and do not begin with a
% lowercase letter or number. Examples of good headings are:

\section*{Keywords/Index Terms}
5464 Remote sensing;
5465 Rings and dust;
6213 Dust;
6250 Moon (1221)

\section{Introduction}
\label{sec:intro}
Lunar dust, because of its adherence and abrasive properties, is prone to damaging spacesuits and instruments, as well as reducing visibility during landing, and thus poses a hazard to lunar exploration \cite{goodwin2002apollo}. Surface electric fields and dust transport on the Moon's surface could also compromise future astronomical observations from the Moon \cite{article}. It is therefore vital to study this dust environment in order to develop solutions for dust-related problems.

Furthermore, characterizing the dust density at different altitudes above the Moon's surface can reveal the amount of dust that gets ejected from the surface and thus clarify dust transport on airless planetary bodies \cite{wang2016dust}.
Dust above airless bodies can be naturally lofted by micro- meteoroid impacts \cite{popel2016impacts}. 
\change{However, very small dust grains (less than about 0.5 $\mu$m in size)  close to the Moon's surface could also be levitated by electrostatic forces.}{ However, some dust grains could also be levitated above the Moon's surface by electrostatic forces} \cite{colwell2009lunar, 2018AdSpR..62..896O}. Furthermore, future missions to the surface can loft significant amounts of dust \cite{metzger2020dust,immer2011apollo,lane2012ballistics} and so understanding the current dust environment can provide a useful baseline for its dynamics and distribution.

Information about the lunar dust exosphere comes both from remote sensing \cite{mccoy1976photometric, 2011P&SS...59.1695G, glenar2014search, feldman2014upper, rennilson1974surveyor}  and in-situ observations \cite{berg1978lunar, grun2011lunar, szalay2015search}. Remote sensing observations include excess brightness measurements in the photometrically calibrated coronal photography and the visual observations of extended lunar horizon glow by Apollo 17 astronauts, which were attributed to forward scattering of sunlight by dust particles \cite{mccoy1976photometric}. A reanalysis of the Apollo observations by \citeA{2011P&SS...59.1695G} concluded that this dust cloud would extend into shadow a distance between 100 and 200 km from the terminator. More recently, faint signals in some of the images from the Clementine navigational star tracker measurements were matched to a dust exosphere with a dust column of about 5 - 30 cm$^{-2}$ and scale heights of 3 km and 12 km. \cite{glenar2014search}. Dust within a few kilometers of the surface is probably undetectable in these measurements since it corresponds to a small part of the pixel's field of view. The Lyman-Alpha Mapping Project far ultraviolet spectrograph on the Lunar Reconnaissance Orbiter (LRO) has estimated upper limits for vertical dust column at less than 10 grains cm$^{-2}$ of 0.1 $\mu$m size \cite{feldman2014upper}. These far-ultraviolet measurements were within the wavelength range 170 and 190 nm and could probe altitudes down to about 10 km, limited by shadowing. Closer to the surface, light from very low-altitude dust was observed by each of the Surveyor 7, 6 and 5 landers along the western lunar horizon following local sunset \cite{rennilson1974surveyor}.

In-situ measurements of lunar dust include data from the Lunar Ejecta and Meteorite (LEAM) Experiment deployed by Apollo 17, which registered a multitude of unexpected hits during lunar sunrise and
sunset, possibly caused by slow moving and highly charged dust grains transported across the lunar surface \cite{berg1978lunar, grun2011lunar}. LEAM events are consistent with the sunrise/sunset–triggered levitation and transport of slow moving, highly charged lunar dust particles \cite{colwell2007lunar} indicating that the amount of lunar dust varies with local time. More recently, the LDEX experiment on board the LADEE spacecraft observed an extremely tenuous asymmetric dust cloud around the Moon with peak densities around 0.004-0.005 m$^{-3}$ near the dawn terminator for a grain size threshold of 0.3 $\mu$m extending hundreds of kilometers above the Moon's surface \cite{horanyi2015permanent}. This instrument also reported an upper limit of 100 m$^{-3}$ on the dust density of particles larger than 0.1 $\mu$m in the altitude range of $\simeq$ 3-250 km for a putative population of electrostatically lofted dust  \cite{szalay2015search}.

In this paper, we analyze the spectra obtained by LADEE's Ultraviolet-Visible Spectrometer (UVS). The UVS made a series of Almost Limb observations from a spacecraft altitude of about 20 km of the lunar sunrise terminator at different locations and times shown in Table~\ref{tab:1}. The observation geometry during the Almost Limb activities enables us to probe the Moon's dust atmosphere at altitudes ranging between 1 and 10 km above the surface. These data probe lower altitudes than any of the prior remote sensing observations from orbiting spacecraft, and even reach below the lowest reported altitudes sensed by the LDEX instrument. Therefore, we aim to constrain whether there is any additional population of dust between 1 and 10 km of the surface beyond the dust cloud seen by LDEX. Given that the nature of the low-altitude dust populations seen by surface observations like LEAM and Surveyor are still uncertain, we will derive empirical limits on this dust population for a range of possible particle populations.

Section~\ref{sec:method} below describes how we processed the Almost Limb data to obtain the tightest limits on potential dust populations. Note that these data contain a fluctuating signal that 
is inconsistent with a dust atmosphere whose density declines with altitude. 
We, therefore, apply a spectral filter to eliminate these fluctuations and isolate signals that can be used to 
set constraints on the dust atmosphere. We then compare these filtered spectra to the expected signals from various dust populations obtained using a Fraunhofer-diffraction model in order to derive upper limits on the dust particle densities. These constraints are summarized in Section~\ref{sec:results} and compared with prior limits on the low-altitude dust atmosphere.

% \section{Materials and Methods}
% Here is text on Materials and Methods.
%

\section{Methods} \label{sec:method}
The process by which we constrain the dust populations above the Moon's surface from the UVS Almost Limb data has multiple steps, which are described in detail below.
Section~\ref{sec:uvsobs} gives a brief description of the UVS instrument and describes in detail the Almost Limb activities and their geometry. Section~\ref{sec:filter} then describes how the spectra are processed to isolate the dust  signal. Section~\ref{sec:model} next discusses how we  produce predicted spectra for various dust densities and particle size distributions using a Fraunhofer diffraction model. Finally, Section~\ref{sec:comparison} describes how we compared our modeled spectra with the spectra obtained by the UVS instrument in order to constrain dust densities above the surface.

\begin{table}[t]
\caption{Almost Limb activities used in this analysis } 
\centering
\begin{tabular}{c c c c c c} 
\hline\hline
\multirow{2}{*}{Filename} & \multirow{2}{*}{Mid time} & \multicolumn{2}{c}{Observed}  & Sub-Solar & Sub-Earth \\ [0.5ex] 
& & latitude &  longitude & longitude & longitude\\
\hline
1836A & 01 Apr 2014 22:40:16 & 24.33$^{\circ}$ & 58.43$^{\circ}$ & 157.88$^{\circ}$ & 5.54 $^{\circ}$\\
1840A & 02 Apr 2014 08:10:22  & 24.19$^{\circ}$ & 53.63$^{\circ}$ & 153.65$^{\circ}$ & 5.58$^{\circ}$\\
1847A & 03 Apr 2014 03:10:28 & 23.82$^{\circ}$ & 44.07$^{\circ}$ & 143.37$^{\circ}$ & 5.48$^{\circ}$\\
1873A & 05 Apr 2014 23:13:22 & 23.00$^{\circ}$ & 9.73$^{\circ}$ & 108.76$^{\circ}$ & 3.53$^{\circ}$ \\
1875A & 06 Apr 2014 04:52:56 & 23.13$^{\circ}$ & 6.83$^{\circ}$ & 105.88$^{\circ}$ & 3.28$^{\circ}$ \\
1880A & 06 Apr 2014 18:05:14 & 23.51$^{\circ}$ & 0.06$^{\circ}$ & 99.17$^{\circ}$ & 2.67$^{\circ}$ \\ 
1882A & 06 Apr 2014 21:51:26 & 23.62$^{\circ}$ & 358.12$^{\circ}$ & 97.25 $^{\circ}$ & 2.48$^{\circ}$\\
1889A & 07 Apr 2014 14:50:09 & 24.16$^{\circ}$ & 349.38$^{\circ}$ & 88.63$^{\circ}$ & 1.63$^{\circ}$\\
1901A & 08 Apr 2014 21:01.12 & 24.94$^{\circ}$ & 333.87$^{\circ}$ & 73.29$^{\circ}$ & 0.04$^{\circ}$\\
1918A & 10 Apr 2014 14:31:13 & 24.86$^{\circ}$ & 317.77$^{\circ}$ & 52.23$^{\circ}$ & -2.05$^{\circ}$\\ 
1929A & 11 Apr 2014 20:42:35 & 25.37$^{\circ}$ & 303.42$^{\circ}$ & 36.91$^{\circ}$ & -3.35$^{\circ}$\\ 
1956A & 14 Apr 2014 17:20:29 & 22.82$^{\circ}$ & 269.43$^{\circ}$ & 2.11$^{\circ}$ & -5.06$^{\circ}$ \\ 
1969A & 16 Apr 2014 11:58:26 & 24.26$^{\circ}$ & 247.64$^{\circ}$ & 340.50$^{\circ}$ & -5.09$^{\circ}$\\ 
1987A & 17 Apr 2014 21:20:19 & 24.85$^{\circ}$ & 230.67$^{\circ}$ & 323.60$^{\circ}$ & -4.57$^{\circ}$\\ 
[1ex]
\hline 
\end{tabular}
\label{tab:1} 
\end{table}

\begin{figure}[t]
\centering
\noindent\includegraphics[width=\textwidth]{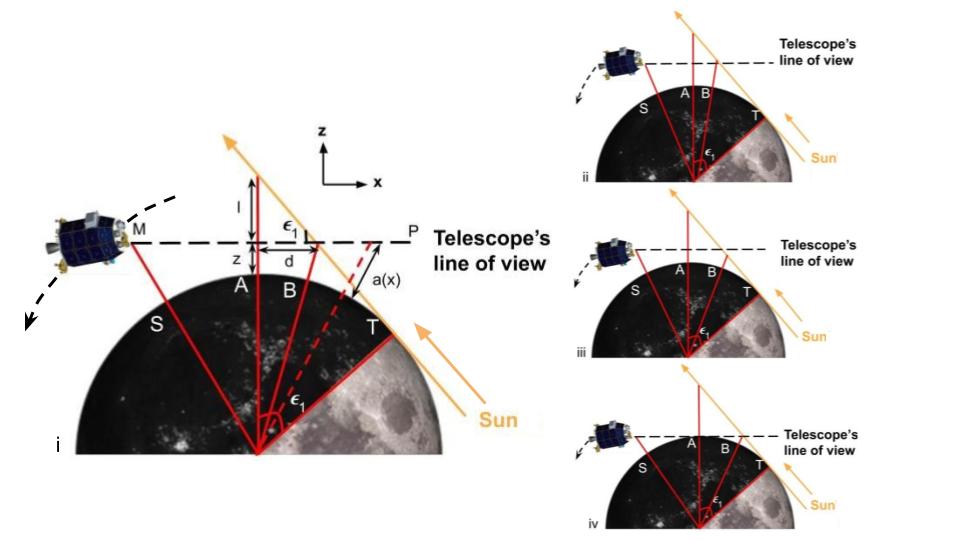}
\caption{Illustrations of the Almost-Limb observation geometry using a 2D projection of the geometry  in the plane containing the UVS, the line of sight and the lunar center.
Note the Moon is not shown to scale in these diagrams. (i) Definitions of relevant geometrical parameters (see also~\ref{appa}). The black dashed line represents the line-of-sight of the telescope, $\epsilon_1$ and $a(x,t)$ are the Sun's elevation angle at that position and the height above the surface for any point lying on the field of view, respectively.  $T$ marks the Terminator point, $S$ is the position on the surface directly below the spacecraft, $A$ is where the telescope's line of sight gets closest to the lunar surface, and $B$ is the point where the line of sight crosses into the shadowed region.  Figures (ii), (iii) and (iv) show how the geometry changes over the course of the Almost Limb activity. During this activity the telescope moves around the Moon, so its line of sight approaches and crosses the limb shortly after the Sun sets. At the end of the Almost Limb activity the telescope field of view, which points at the fixed direction relative to the Sun, intercepts the Moon's unlit surface.}
\label{fig:1}
\end{figure}

% \subsection{A descriptive heading about methods}
% More about Methods.
%

\subsection{UVS Almost Limb observations}
\label{sec:uvsobs}
The LADEE-UVS instrument is described in detail in \citeA{2014SSRv..185...63C}, but for the sake of completeness we review some of its properties here.

This instrument observes light with wavelengths between 250 and 800 nm over a 1-degree field of view via a catadioptic telescope that is coupled by a fiber-optic cable to a spectrometer with a spectral resolution of $\lambda/\Delta\lambda \sim$ 900 at 500 nm.
The spectrometer disperses this light onto a 1044$\times$64 (1024$\times$58 active) pixel detector CCD array. Each column of pixels is binned within the detector, delivering a 1$\times$1044 pixel spectrum. The raw signals recorded by the instrument are processed by standard pipelines in four steps: dark and bias correction, normalization by integration time, second order grating effects correction, and finally the radiance calibration. Each calibrated spectrum recorded by LADEE-UVS therefore contains 1024 active records of radiance
in units of W m$^{-2}$ nm$^{-1}$ sr$^{-1}$. However, when considering the light scattered by solid material like dust or the Moon’s surface, it is more useful to instead consider reflectance, which is typically expressed in terms of the unitless ratio $I/F$, where $I$ is the intensity of the scattered radiation and $F$ is the solar flux density (solar flux divided by $\pi$). Hence for this investigation we converted the UVS radiance data to $I/F$ values using a standard solar spectrum \cite{mecherikunne} at 1024 wavelengths ranging from 229.26 nm to 812.55 nm.

The LADEE mission executed science observations in lunar orbit spanning 2013 Oct 16 - 2014 Apr 18. Over the duration of the LADEE mission, UVS executed 1890 activities, collecting over 1 million spectra. During the last month of the mission, the UVS instrument obtained 15 so-called ``Almost Limb'' observations as the spacecraft lowered its altitude before crashing into the surface. These observations will be the exclusive focus of this study.

The 14 Almost Limb activities included in this analysis are summarized in Table~\ref{tab:1} (One Almost Limb activity, designated 1855A, did not contain any useful data and is therefore omitted from this analysis).
The duration for each of these activities was approximately 8 minutes. Table~\ref{tab:1} also shows the latitude and longitude of the point where the telescope's line of sight first intersected the Moon's surface during the course of each observation.

During each Almost Limb activity, the spacecraft flies over the Moon on a retrograde orbit at an approximately constant altitude while the telescope points in a fixed direction relative to the Sun, so that the telescope stares at a fixed location in the sky 12$^\circ$ away from the Sun that crosses the limb shortly after the Sun sets. Figure~\ref{fig:1} is a 2D projection of the observation geometry in a plane containing UVS line of sight and the Lunar center. This type of observation allows the faint signals from dust near the Moon's surface to be cleanly isolated from other astronomical signals like zodiacal light. 

Zodiacal light is a diffuse signal created by sunlight scattered off of interplanetary dust \cite{1998A&AS..127....1L,article,2020P&SS..19004973L} that is often a significant background for remote-sensing searches for lunar dust \cite{stubbs2010optical,feldman2014upper}. The brightness of zodiacal light relative to Lunar Dust is expected to be lower at the ultra-violet wavelengths observed by LADEE-UVS than it is at visible wavelengths \cite{stubbs2010optical}, but it cannot be completely  neglected. Fortunately, during each Almost Limb activity LADEE-UVS stares at a single point in the sky that is nearly fixed relative to both the Sun and distant stars (the apparent motion of the Sun being negligible over the few minute duration of a typical observation). The signals from zodiacal light and other astronomical backgrounds therefore should remain constant over the course of each observation. By contrast, the signal from dust should increase as the line-of-sight approaches the Moon's surface, producing a time-variable signal that can be cleanly separated from such constant backgrounds.

\begin{figure}[ht]
\centering
\noindent\includegraphics[width=\textwidth]{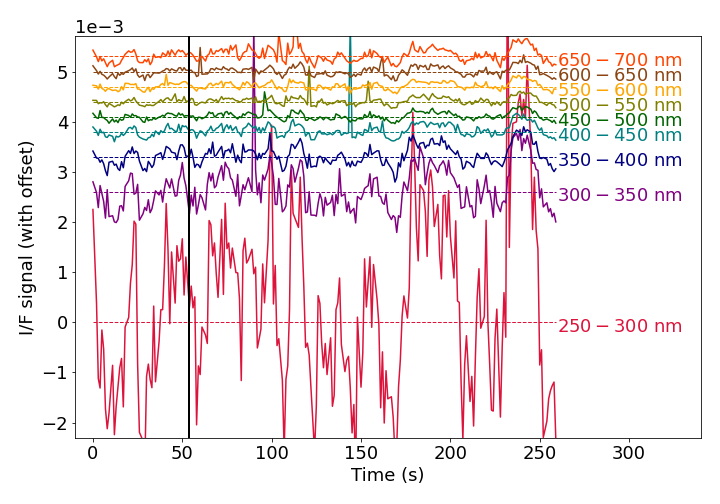}
\caption{Unfiltered $I/F$ values for different wavelength ranges as recorded by the UVS instrument on board LADEE during the Almost Limb activity, 1969A. An offset has been added to these values to view the various wavelength ranges distinctly. The vertical black line on the plot denotes the point ($t_V$) when the telescope's line of view hits the Moon's surface. The fluctuations in the signal level persist even after the telescope's field of view crosses the Moon's terminator, indicating that these fluctuations are likely instrumental artifacts.}
\label{fig:2}
\end{figure}

\subsection{Spectral filtering} \label{sec:filter}

Initial looks at the LADEE-UVS Almost Limb data revealed signal fluctuations that were inconsistent with the model of a uniform dust atmosphere. 
The signal from dust is expected to be broadband \cite{stubbs2010optical,van1957light} so we calculated the average of reflectance values for wavelength ranges of width 50 nm from 250-300 nm to 650-700 nm. (Wavelengths shorter/longer than the above range have not been considered for this analysis because their signal-to-noise ratio was poor.) Figure~\ref{fig:2}  shows the resulting average $I/F$ values for the Almost Limb activity 1969A after the Sun sets. The vertical black line on the plot, denotes the point ($t_V$) when the telescope's line of view meets the Moon's surface. An offset has been added to the spectral channels to view each wavelength signal distinctively.

The signals in Figure~\ref{fig:2} do not show any obvious trends with altitude above the limb that would be expected from exospheric dust. Instead, the signals contain fluctuations 
that persist beyond the point when the telescope is viewing the unlit part of the lunar surface. This is not consistent with any dust atmosphere, which would produce a signal that increases with decreasing altitude above the limb and then disappear when the telescope starts viewing the unlit surface. 

The origins of these fluctuations are still unknown. The signal fluctuations seen in Figure~\ref{fig:2} have a similar shape in all the frequency bands, so they likely have a common source with a distinct broad-band spectrum. However, since the spacecraft is in shadow and viewing the dark side of the Moon for part of the time the spacecraft sees these signals (cf. Figure~\ref{fig:1}), it seems unlikely that they represent fluctuations in the amount of light entering the instrument. Likewise, these signals also do not show any correlation with variations in instrument parameters like the target and detector temperature, so they cannot be clearly associated with something internal to the instrument.

Regardless of their origin, these fluctuations obscure the signal from the dust atmosphere. Fortunately, these variations have a different spectrum from dust, and so we can use spectral filtering techniques to remove these variations and isolate potential dust signals in order to constrain any real signals from low-altitude dust.

\begin{figure}[t]
\centering
\noindent\includegraphics[width=\textwidth]{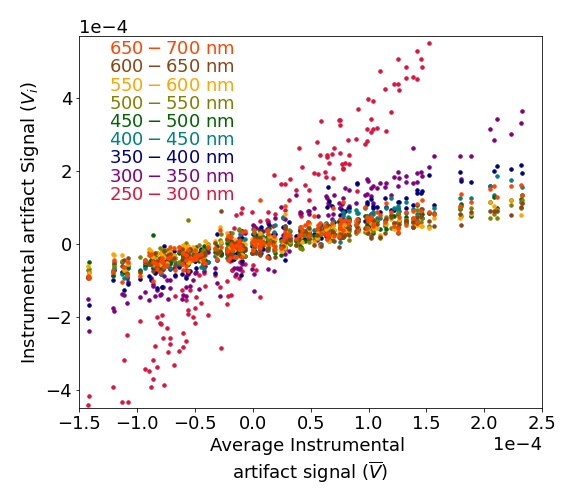}
\caption{The varying signal, $V_i(t>t_V)$ for each wavelength band in the Almost Limb activity 1969A plotted against the average signal, $\bar{V}(t>t_V)$ over the whole wavelength range (250-700 $nm$). Note the strong correlations between each individual spectral channel and the average signal.}
\label{fig:3}
\end{figure}

\begin{figure}[ht]
\centering
\noindent\includegraphics[width=\textwidth]{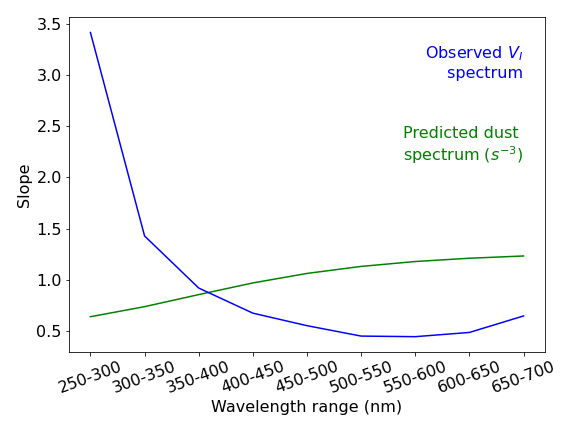}
\caption{The slope $m_{V_i}$ from equation~\ref{eq:2} for a linear fit of the scatter points in Figure~\ref{fig:3} between the variance signals at each wavelength $V_i(t>t_V)$ versus the average variance signal $\bar{V}(t>t_V)$ for the Almost Limb activity 1969A is shown as the blue line. The green line for comparison, is the predicted slope for a dust atmosphere with a power-law particle size distribution ($s^{-3}$ with a threshold of 0.3 $\mu m$) described in Section~\ref{sec:model}. These slope curves are equivalent to normalized spectra of the varying signal and dust, respectively. Since these two spectra are so different, filtering out a signal with the blue spectrum will not eliminate signals with something like the green spectrum.}
\label{fig:4}
\end{figure}

Let us assume that the total signal $S_i$ in a wavelength channel $i$ in Figure~\ref{fig:2} is a combination of a dust signal $D_i$ and the varying signal $V_i$, both of which change with time such that, 
\begin{equation}
S_i(t) = D_i(t) + V_i(t).
\label{eq:1}
\end{equation}
The key thing to note is that any spectrum beyond the point of time, $t_V$ at which the line of view hits the surface is entirely due to $V_i(t)$ such that, $S_i(t>t_V)$ = $V_i(t>t_V)$. 
If the fluctuating signal has a fixed spectrum, then the value of $V_i(t>t_V)$ at any single wavelength should be a linear function of the average value of $V_i(t>t_V)$ over all wavelengths (i.e. $\bar{V}(t>t_V)$ = $\sum_i V_i(t>t_V)/N$, where $N$ is the number of wavelength bands). Figure~\ref{fig:3} shows the varying background, $V_i(t>t_V)$ for each wavelength range, $i$ plotted against the average background, $\bar{V}(t>t_V)$ for the complete wavelength range (250-700 nm) for the Almost Limb activity 1969A and considering only times $t>t_V$. This plot confirms that the signal at each wavelength is indeed proportional to the average signal over all wavelengths.
We therefore perform a linear fit to the values of $V_i(t>t_V)$ in each spectral channel $i$ as a function of the mean signal $\bar{V}(t>t_V)$ and use the slope and offset of these fits ($m_{V_i}$ and $b_{V_i}$) to derive the following estimates of $V_i(t)$ at each wavelength:

\begin{equation}
V_{i,est}(t) = m_{V_i}\bar{V}(t) + b_{V_i}.
\label{eq:2}
\end{equation}

Figure~\ref{fig:4} shows the derived values of $m_{V_i}$ as a function of wavelength, which is essentially the normalized spectrum of $V_i$.  The spectrum of this signal is clearly different from the spectrum of the expected dust signal (derived in Section~\ref{sec:model}), so it should be possible to remove $V_i$ without completely removing the dust signal. Specifically, consider the following quantity:
\begin{equation}
D_{i,est}(t) = S_i(t) - m_{V_i}\bar{S}(t) - b_{V_i}
\label{eq:3}
\end{equation}
where $\bar{S}=\sum_i S_i(t)/N$, and $m_{V_i}$ and $b_{V_i}$ are the same slope and intercept parameters from Equation~\ref{eq:2} derived from the data obtained at times $t>t_V$. If we insert $S_i(t) = D_i(t) + V_i(t)$ into this expression, we obtain:
\begin{equation}
D_{i,est}(t) = D_i(t) + V_i(t) - m_{V_i}\bar{D}(t) - m_{V_i}\bar{V}(t) - b_{V_i}
\label{eq:4}
\end{equation}
where $\bar{D}(t)=\sum_i D_i(t)/N$.
Provided that $V_{i,est}(t)$ is a good estimate of $V_i(t)$ both before and after $t_V$, the terms involving $V_i(t)$ and $\bar{V}(t)$ will cancel, leaving
\begin{equation}
D_{i,est}(t) = D_i(t) - m_{V_i}\bar{D}(t).
\label{eq:5}
\end{equation}
 The quantity $D_{i,est}$ should therefore not contain any contamination from the fluctuating instrument signal, at the cost of the dust signal being attenuated by a predictable factor. (In practice, we also need to remove a small number of data points that are large outliers to the mean trends, which we do by excluding any data where $S_i(t)$ is more than 3$\sigma$ from its mean value in any of the wavelength bands.)

\begin{figure}[t]
\centering
\noindent\includegraphics[width=\textwidth]{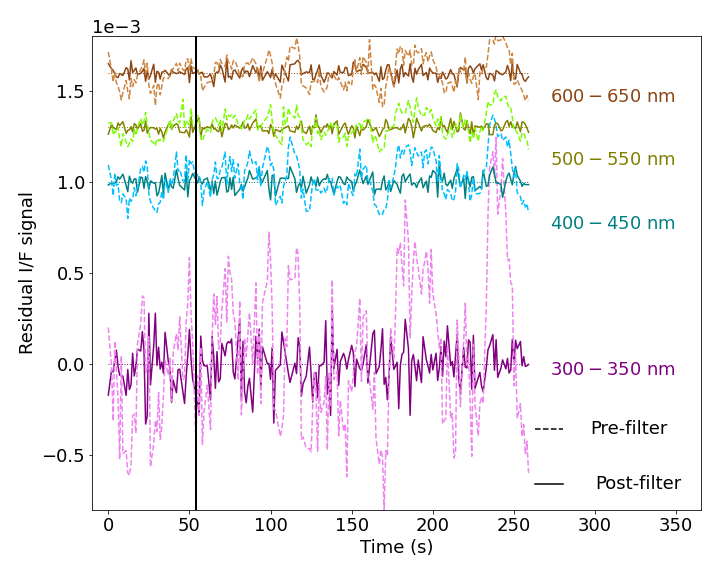}
\caption{Corrected signal, $D_{i,est}(t)$ for the  Almost Limb activity 1969A, after spectral filtering has been applied to each wavelength range (solid lines) compared with the original data (dashed lines). The vertical black line on the plot, denotes the time ($t_V$) when the telescope's line of view hits the Moon's surface. An offset has been added here too to view each wavelength range distinctly. Note the filtered data shows much smaller fluctuations than the raw data.}
\label{fig:5}
\end{figure}
 
 The filtered signal $D_{i,est}(t)$ for observation 1969A is shown in Figure~\ref{fig:5} and indeed the fluctuations are much reduced, as desired. The results of applying this filtering technique to the other Almost Limb activities are shown in~\ref{appn2}. In all cases, the fluctuations are significantly reduced. 
\begin{figure}[t]
\centering
\noindent\includegraphics[scale=0.6]{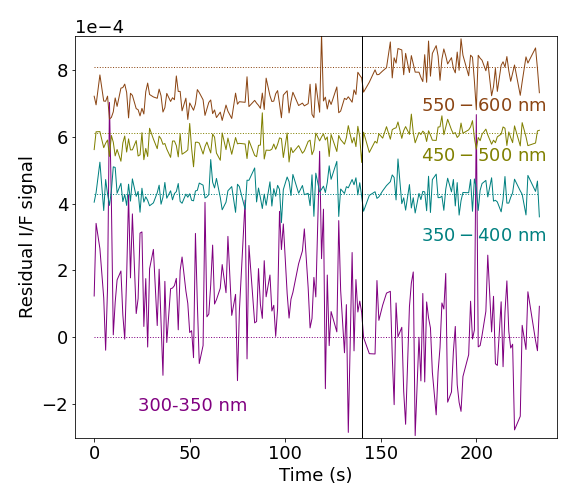}
\caption{Corrected signal, $D_{i,est}$ for the Almost Limb activity 1840A, after spectral filtering has been applied to each wavelength range. The vertical black line on the plot, denotes the time ($t_V$) when the telescope's line of view hits the Moon's surface. An offset has been added here too to view each wavelength range distinctly. Note the slight shifts in the signal level beyond with the vertical line. These shifts are likely due to Earthshine from the Moon's surface, which is expected to be present in these early observations. The wavelength trends observed here are also consistent with the red color expected for Earthshine.}
\label{fig:6}
\end{figure}
 
 When we apply this filtering technique to the first Almost Limb activities, we observe an interesting phenomenon that further illustrates this technique can reveal real signals in the data. Figure~\ref{fig:6} shows an example of the results from processing one of these earlier observations (1840A). In this case, the filtered signal shows a clear brightness change at the limb crossing time $t_V$, with the signals at short wavelengths becoming more negative while those at long wavelengths become more positive. Since these signals do not change with altitude above the limb, these are probably not real dust signals. Instead, they are almost certainly due to Earthshine, which causes the surface of the Moon to be slightly brighter than the background sky. Earthshine is expected to have a red color, which is consistent with the wavelength trends observed here. Furthermore, these shifts are only clearly visible in the earliest observations, which are also the ones expected to be most strongly affected by Earthshine (see Table~\ref{tab:1} and Figure~\ref{fig:20}). This demonstrates that our spectral-filtering process is able to reveal real astronomical signals that were previously obscured by the instrumental artifacts, and so means that we can also use these filtered data to search for signals from lunar dust. Taking a closer look at the other observations reveals that 9 observations have significant contamination from Earthshine. But the remaining 5 observations can still be used to search for dust (see Section~\ref{sec:comparison} below).

\subsection{Predicted signals from model dust atmospheres}
\label{sec:model}
Since the filtered signals do not show clear variations with altitude that could be due to lunar dust, we need an explicit model to constrain the amount of dust that could be present. If we assume a density of particles at each altitude, the spectra can be computed using the appropriate scattering theory. For a flux of radiation $\pi F$ with wavelength $\lambda$, the power scattered by an individual particle per unit solid angle is given by the function $\frac{dP}{d\Omega}$. %This function depends on $F$, $\lambda$, scattering angle $\theta$, the radius of the particle $s$, and the optical properties of the particle. 
In principle, these spectra could be computed using Mie Theory or even more sophisticated scattering theories that can account for the irregular shapes of these particles \cite{shkuratov1994critical,Kolokolova2015pol}. However, in practice the Almost Limb activities were made at such high phase angles (168.3$^{\circ}$) that the observed light is predominantly due to diffraction around individual particles, and the lunar dust populations are sufficiently tenuous that multiple scattering among different particles can be ignored. In this situation, Fraunhofer diffraction theory therefore provides a more efficient but still sufficiently accurate way to estimate the spectra of the relevant dust populations.

The power scattered per unit solid angle by a conducting disk of radius $s$ illuminated by flux $\pi F$ of radiation with wavelength $\lambda$ is given by the following function \cite{jackson1975multipoles, 2009ApJ...693.1749H}:
\begin{equation}
\frac{dP}{d\Omega}=\pi F s^2\frac{ J_1^2(k\, s\, \sin\theta)}{\sin^2\theta}
\label{eq:6}
\end{equation}
where $s$ is the size (radius) of the particle, $k$=2$\pi$/$\lambda$, $J_1$ is the spherical Bessel function of first kind and $\theta$ is the scattering angle ($180^\circ$-the phase angle).
The above expression holds true for conducting materials, for dielectric particles we need to multiply the above expression by a scaling factor \cite{fymat1981mie}. This scaling factor can be derived directly from the extinction factor $Q_{ext}$ (the ratio of the total cross-section of the particle to its geometrical cross-section, see~\ref{appa}) using the optical theorem which gives the following approximation for the scattering by a dielectric sphere:
\begin{equation}
\frac{dP}{d\Omega}=\frac{\pi F s^2}{4\sin^2\theta} J_1^2(k\,s\,\sin\theta) Q_{ext}^2(k\,s,m).
\label{eq:7}
\end{equation}
The above equation defines the power scattered per unit solid angle for particles of size $s$. This value of power scattered obtained using the Fraunhofer model differs from the Mie theory values by only a factor of about 0.97 - 1.8 for the ranges of particle sizes and observing geometries considered here, and hence can be considered a reasonable approximation of the expected signals from spherical particles. \change{Furthermore, at low scattering angles (high phase angles) the signals from compact irregular grains are similar to those from spheres with equivalent volume}{ Furthermore, at low scattering angles (high phase angles) the signals from compact irregular grains differ from spheres with equivalent volume only by atmost 50$\%$.} \cite{pollack1980scattering}. Our Fraunhofer calculations therefore also provide reasonable approximations for the signals for irregular grains with the equivalent effective sizes $s$.

The dust above the Moon's surface is a collection of particles of different size ranges for which a size distribution can be defined. The reflectance of this collection of particles in a given wavelength channel $i$ and time $t$ is given by the standard unitless quantity:
\begin{equation}
\Bigg[\frac{I}{F}(t)\Bigg]_{i,pred}=\frac{1}{F}\int_{s_{min}}^{s_{max}} \frac{dP_i}{d\Omega}\mathcal{N}(s,t)ds.
\label{eq:8}
\end{equation}
where $\frac{dP_i}{d\Omega}$ is the scattered power in wavelength channel $i$ and $\mathcal{N}(s,t)$ is the particle size distribution integrated along the line of sight during that particular time. The limits on the above integral are defined by the lower and upper limits of the size distribution discussed below.

For this analysis, we will consider several different models for the spatial and size distributions of particles. However, we will assume that the particle size distribution is independent of location, and only the total density of particles varies with altitude above the Moon's surface. These assumptions not only make the calculations more tractable by restricting the phase space of possible predictions, but also facilitate comparisons with previously published measurements of these dust populations.

We consider two different types of particle size distributions in this study.
On the one hand, we consider narrow size distributions for particles with radii between 0.07 to 1 $\mu$m in order to illustrate how sensitive these observations are to particles of different sizes \cite{feldman2014upper, glenar2014search}. Also, narrow size distributions could be more representative of electrostatically charged dust grains \cite{criswell1973horizon,rennilson1974surveyor}. 

On the other hand, we consider power-law size distributions because these size distributions are also more likely to be representative of material lofted by impacts, which are expected to have a broad size distribution \cite{horanyi2015permanent,grun2011lunar}. More specifically, we will use power-law size distributions with limits at 0.3 $\mu$m and 10 $\mu$m. These limits have been chosen to facilitate comparisons with previous limits on dust density \cite{horanyi2015permanent,szalay2015search,glenar2014search,feldman2014upper} and because UVS operates at wavelengths between 200  and 800 nm and these sorts of remote sensing observations are more sensitive to particles comparable to the operating wavelength range of the instrument \cite{van1957light}.
 Also, we consider two different cases for the power-law index:  $\mathcal{N}(s,t)\propto s^{-3}$ and  $\mathcal{N}(s,t)\propto s^{-4}$. These two numbers (-3 and -4) bracket the expected differential size distribution of collision debris, which has a differential index of around -3.5 \cite{dohnanyi1969collisional,tanaka1996steady}, and the value of -3.7 measured in-situ by LDEX  \cite{szalay2016lunar}.

In addition to assuming a specific size distribution, we also need to assume the dust has a specific spatial distribution before we can compute the predicted signals and how they vary with time.
Since this analysis seeks to constrain a low-altitude global dust atmosphere, we will assume that the number of particles per unit area along the line of sight can be described by a simple exponential dust profile defined by nominal total number density of particles at the surface $n_{o,nom}$ and scale height $H$. \change{Such models are almost certainly an oversimplification because the actual dust distribution is a convolution of the launch velocity distribution of the particle populations ejected from the surface, which depends on how the particles are ejected, as well as their size and charge state.}{ Such models are almost certainly an oversimplification because the actual dust distribution from both electrostatically lofted and impact-generated dust is a convolution of the launch velocity distribution of the particle populations ejected from the surface, which depends on how the particles are ejected, as well as their size and charge state. Both methods of launching dust can potentially send material to a wide range of altitudes} \cite{szalay2015search,horanyi2015permanent}. However, a simple exponential profile is a reasonable choice for this initial study, primarily because it facilitates comparisons with prior remote-sensing searches for and in-situ measurements of lunar dust, which have assumed exponential density profiles \cite{feldman2014upper, glenar2014search, szalay2016lunar}. 
In this case, the total number of particles per unit area along the line of sight is given by the integral
\begin{equation}
N(t)=n_{o,nom}\int_{d}^{\infty}e^{-a(x,t)/H}dx
\label{eq:9}
\end{equation}
where $x$ denotes the distance along the line of sight and $a(x,t)$ is the altitude above the surface at a particular time (see Figure~\ref{fig:1}). 
This integral is evaluated from  a lower limit $d$ that signifies the point where the dust is illuminated to an upper limit of $+\infty$ (see Figure~\ref{fig:1}). The lower limit $d$ and $a(x,t)$ 
are both explicitly calculated in~\ref{appa}. 
 For this study, we will specifically consider scale heights $H$ of 1, 3 and 5 km. These values were chosen because the LADEE-UVS data are most sensitive to variations in the dust density on these scales. These also correspond to scales that have not yet been probed by previous remote-sensing observations \cite{feldman2014upper, glenar2014search} and are near the lower limits of the published LDEX observations \cite{horanyi2015permanent, szalay2015search}.

Finally, in order to account for the spectral filtering of the observations, we apply the same filter to the predicted $I/F$ signal as we did to the observed data (see Equation~\ref{eq:3}) to create the following quantity.
\begin{equation}
P_{i,est}(t) = \Big[\frac{I}{F}(t)\Big]_{i,pred} - m_{V_i}\overline{\Big[\frac{I}{F}(t)\Big]}_{pred} - b_{V_i},
\label{eq:10}
\end{equation}
where $I/F_{i,pred}$ is the predicted $I/F$ for each wavelength channel $i$, $\bar{I/F}_{pred}$ is the average predicted $I/F$ over all wavelengths, and $m_{V_i}$ and $b_{V_i}$ are the same parameters used in Equation~\ref{eq:3}.
The resulting estimates of $P_{i,est}(t)$ can therefore be directly compared to the filtered observed values of $D_{i,est}(t)$ to derive limits on the dust signals.

\subsection{Comparison of UVS spectra and the Predicted signal}
\label{sec:comparison}

\begin{figure}[ht]
\centering
\noindent\includegraphics[width=\textwidth]{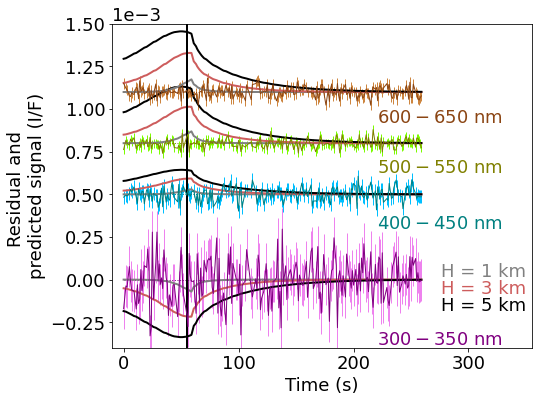}
\caption{This figure shows the filtered residual $I/F$ signal, $D_{i,est}(t)$ with error bars for Almost Limb activity 1969A at each wavelength range. The error bars are based on the scatter in the data points. The vertical black line on the plot, denotes the point ($t_V$) when the telescope's line of view hits the Moon's surface. The signal, $P_{i,est}(t)$ predicted using the Fraunhofer model and the exponential density profile for a size distribution proportional to $s^{-3}$ and for scale heights of 1, 3 and 5 km, is plotted over the residual signal in grey, light coral and black, respectively. An offset has been added to view the signals distinctly. Note that the predicted signal, $P_{i,est}(t)$ is estimated for a nominal surface concentration of $n_{o,nom} = 10^3 m^{-3}$ and is greater than the residual signal, indicating that the upper limit on dust density will be below this value.}
\label{fig:7}
\end{figure}

\begin{figure}[t]
\centering
\noindent\includegraphics[width=\textwidth]{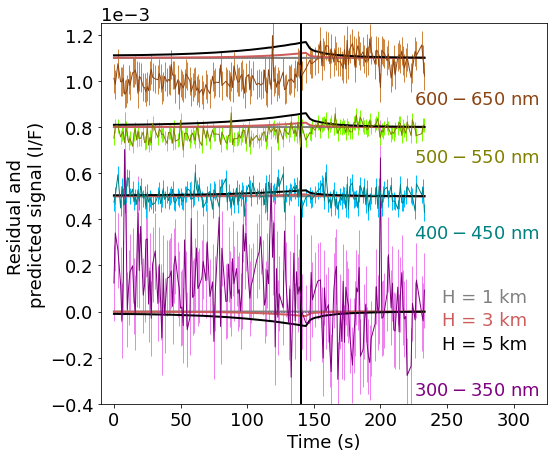}
\caption{This figure shows the filtered residual $I/F$ signal, $D_{i,est}(t)$ with error bars for Almost Limb activity 1840A at each wavelength range. The error bars are based on scatter in the data points. The vertical black line on the plot, denotes the point ($t_V$) when the telescope's line of view hits the Moon's surface. The signal predicted using the Fraunhofer model and the exponential density profile for a size distribution proportional to $s^{-3}$, nominal surface dust densities of $10^3$ $m^{-3}$ and for scale heights of 1, 3 and 5 km is plotted over the residual signal in grey, light coral and black, respectively. An offset has been added to view the signals distinctly. This plot is an example of the Almost Limb activity contaminated with Earthshine and hence is not included in calculating the upper limit on dust density.}
\label{fig:8}
\end{figure}

Figures~\ref{fig:7} -~\ref{fig:8} show the predicted signal, $P_{i,est}(t)$ for a nominal surface concentration of $10^3$ $m^{-3}$ and scale heights of 1, 3 and 5 km overlaid on the observational data, $D_{i,est}(t)$ from Figures~\ref{fig:5} and~\ref{fig:6} with error bars based on the scatter in the data points (Figures~\ref{fig:13},~\ref{fig:15},~\ref{fig:17} and~\ref{fig:19} for Almost Limb activities 1918A-1987A can be found in the~\ref{appn2}). As expected, the predicted signals become stronger as the telescope's line-of-sight approaches the lunar surface. Note that at short wavelengths the predicted signals are negative because $m_{V_i}$ becomes large enough to flip the sign of the difference in Equation~\ref{eq:10}.

In Figure~\ref{fig:7} the observed brightness trends are much smaller than the predicted trends for dust populations with an arbitrarily chosen nominal dust density of $n_{o,nom}=10^3$ $m^{-3}$. The limits these data can place on these low-altitude dust populations are therefore well below the value of $n_{o,nom}$. We derive quantitative limits on the dust density from each spectral channel by performing a least-squared linear fit of the observed signal calculated in equation~\ref{eq:5} versus the model predictions from equation~\ref{eq:10}:
\begin{equation}
D_{i,est}=\frac{n_{o,est}}{n_{o,nom}}{P_{i,est}}+C.
\label{eq:11}
\end{equation}
Note the slope of this fit is simply the ratio of the most likely value of the surface dust density $n_{o,est}$ to the nominal surface dust density assumed in computing $P_{i,est}$.
We also take the weighted average of these slope and offset estimates from the different wavelength channels for each observation to obtain more precise estimates on these parameters, and propagate errors appropriately.

\begin{table}[ht]
\label{tab:2} 
\caption{Chi squared values and probabilities to exceed for the last five data sets for the size distribution $s^{-3}$ and scale heights $H$ = 1, 3 and 5 km.}
\centering
\begin{tabular}{c| cc | cc| cc} 
\hline
\multirow{2}{*}{Activity name} & \multicolumn{2}{c|}{H = 1 km} & \multicolumn{2}{c|}{H = 3 km} & \multicolumn{2}{c}{H = 5 km}\\
\cline{2-7}
&Chi-square& Probability & Chi-square & Probability & Chi-square & Probability\\
\hline
1918A & 0.05 & 82 & 0.75 & 39 & 1.18 & 28 \\
1929A & 0.13 & 72 & 0.41 & 52 & 1.03 & 31\\
1956A & 0.06 & 81 & 0.04 & 84 & 0.02 & 89\\
1969A &  0.42 & 52 & 0.08 & 78 & 0.38 & 54\\
1987A & 1.82 & 18 & 1.30 & 25 & 1.31 & 25\\  [1ex]
\hline 
\end{tabular}
\end{table}

In principle, the offset $C$ in the above equation should be equal to zero so long as there are no other sources of signal in the data. However, in practice we find that Earthshine contaminates many of these observations, making them less suitable for deriving tight constraints on the dust populations. Figure~\ref{fig:8} shows a clear example of this contamination. Since our filtering method forces the signal to be zero after the vertical line $t_V$, the filtering step causes significant offsets in the signals observed at earlier times. These signals do not vary with altitude like a low-altitude dust atmosphere signal should and so produce significant offsets in the trends of observed versus predicted signals (see Table~\ref{tab:5} in\ref{appn3}). We force this offset $C$ to zero for our estimates of dust density limit. 

 Fortunately, as New Earth approaches and the Earth-Moon phase angle increases, the illumination due to Earthshine on the surface of the Moon decreases. The corresponding values of Earth-Moon system are shown in the Table~\ref{tab:4} and Figure~\ref{fig:20} in~\ref{appn3}. The last three out of the 14 Almost Limb activities (1956A-1987A) are located beyond the Earthshine horizon. In addition, the preceding two Almost Limb activities (1918A and 1929A) were recorded at locations that have high Earth-Moon phase angle (129$^{\circ}$ and 141$^{\circ}$ respectively) and should receive negligible amounts of Earthshine. 
 
We verify that Earthshine is negligible for the last five observations using the above linear model by checking that the wavelength-averaged value of $C$ was consistent with zero for all of these five observations (see Table~\ref{tab:5} in~\ref{appn3}). We also computed the chi-squared statistics of a model where $C$ was forced to be zero for all wavelengths for each of the observations and verified that the resulting individual estimates for $n_{o,est}$ from the last five measurements are all consistent with zero (see Table~\ref{tab:2}). This demonstrates that for these five observations there is no evidence for extraneous signals, and so we can use these five observations to place firm constraints on the low-altitude dust population. Note that these last few activities also probe the lowest altitudes, so they provide the tightest limits on the dust populations at low altitudes.

\section{Results and Discussion}
\label{sec:results}

Our final constraints on low-altitude dust populations are based on the weighted average of the estimates of $n_{o,est}$ from the Almost-Limb observations with negligible Earthshine contamination. These weighted averages are all comparable to their corresponding uncertainties, and so these constraints will mostly be provided as upper limits. These upper limits depend on both the assumed scale height and particle size distribution, and so these different assumptions will be considered separately below.

\begin{figure}[t]
\centering
\noindent\includegraphics[width=\textwidth]{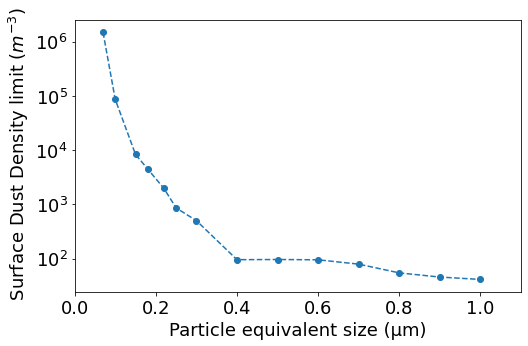}
\caption{The upper limits on surface dust density, $n_{o,est}$ for mono-disperse size distribution with dust particles spherical-equivalent size (see Section 2.3) ranging from 0.07 to 1 $\mu$m derived for the 5 Almost-Limb observations in Table~\ref{tab:2} with negligible Earthshine. 
The scale height $H$ for all these values is 1 km. These limits are within the 95$\%$ confidence level, assuming a positive density value. In principle, this mono-disperse size distribution can be convolved to get upper limits for an arbitrary size distribution. }
\label{fig:9}
\end{figure}

First, consider the limits derived assuming mono-disperse size distributions, since these reveal how our constraints depend on the assumed spherical-equivalent particle size. Figure~\ref{fig:9} shows the 95\% confidence interval limit on dust populations with scale height $H$ of 1 km for mono-disperse size distributions. These limits tighten dramatically with increasing particle size up to around 0.3 $\mu$m, at which point they become much less sensitive to the assumed particle size. This demonstrates that the LADEE-UVS data provide the tightest constrains on the dust density for particles with equivalent radii bigger than 0.3 $\mu$m. This is reasonable, given the UVS data used in this analysis was obtained at wavelengths between 200 nm and 800 nm, and light is most efficiently scattered by particles that are at least as large as the light's wavelength \cite{van1957light}.

\begin{figure}[t]
\centering
\noindent\includegraphics[width=\textwidth]{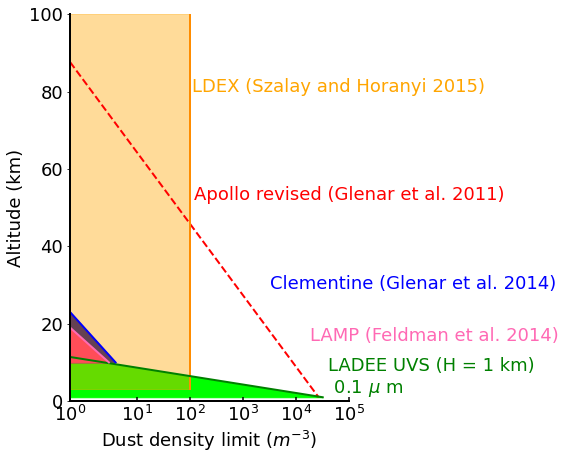}
\caption{A comparison of the upper limits on dust density for particle size 0.1 $\mu$m as a function of altitude obtained from different measurements. This figure uses values for earlier data summarized in \citeA{szalay2015search}. The LADEE-UVS observation plotted here were computed using the last five Almost Limb activities which have low Earthshine signal and corresponds to a scale height $H$ of 1 km. This value of dust density was calculated using a mono-disperse size distribution for a particle of size $0.1 \mu m$ to facilitate comparison with previous limits. Note that the upper limit from the LADEE-UVS observations extend closer to the surface of Moon than prior limits.}
\label{fig:10}
\end{figure}

Many of the prior works on Lunar dust reported limits on particles around 0.1 $\mu$m in size, which is also a reasonable size for a potential population of electrostatically-lofted grains.  We therefore compare these prior observations with our corresponding limit for a mono-dispersed particle size distribution in Figure~\ref{fig:10}.
The specific model limit shown in this figure correspond to a mono-disperse size distribution for an equivalent  particle size of 0.1 $\mu$m and a scale height of 1 km, for which our upper limit on the surface dust density, $n_{o,est}$ is 3.1 $\times 10^4$ $m^{-3}$. The diagonal line shows the corresponding limit on the dust density as a function of altitude for this particular limit, and the green shaded area shows the allowed range of dust densities for this particular model.

Our limit on the surface dust density is of the same order of the estimates of surface dust density derived from Apollo 15 coronal photographs \cite{2011P&SS...59.1695G}, and is several orders of magnitude higher than limits set by more recent remote-sensing measurements \cite{feldman2014upper, glenar2014search}. However, those measurements were insensitive to dust below altitudes of 10 km,  
and so our limit extends to lower altitudes. Furthermore, because we can consider much lower scale heights with the UVS data, at 10 km our limits become comparable to those earlier remote-sensing limits.
Meanwhile, the LDEX instrument reported limits of order 100 $m^{-3}$ in this same size range at altitudes between 3 km and 250 km \cite{szalay2015search}. Our limit on the dust density at 3 km (1500 $m^{-3}$) is an order of magnitude above the LDEX limit. However, our constraints also apply to altitudes below 3 km (and even below 1 km), where LDEX was not able to observe.

\begin{table}[ht]
\caption{Dust density limits $n_{o,est}$ with their respective $1\sigma$ error bars.} 
\label{tab:3} 
\centering
\begin{tabular}{c|cc|cc}
\hline\hline
\multirow{3}{*}{Scale Height
H (km)} & \multicolumn{4}{c}{Dust Density ($m^{-3})$}\\
 \cline{2-5}
 & \multicolumn{2}{c|}{1956A-1987A} &
 \multicolumn{2}{c}{1918A-1987A} \\
 \cline{2-5}
 & $s^{-3}$ & $s^{-4}$ & $s^{-3}$ & $s^{-4}$ \\ [0.5ex] 
\hline 
1 & 75.6$\pm$66.9 & 94.5$\pm$93.7 & 64.6$\pm$63.4 & 83.6$\pm$88.7 \\
3 & $4.6\pm10.6$ & $3.3\pm14.8$ & 8.2$\pm$9.3 & 9.6$\pm$12.9  \\ 
5 & $2.6\pm6.1$ & $2.1\pm8.6$  & 6.0$\pm$5.1 & 7.6$\pm$7.1  \\ 
[1ex]
\hline 
\end{tabular}

Error bars are defined as $1/(\sqrt{(\sum{W_i})})$ where the weights $W_i$ are the errors corresponding to each observation. These upper limits on surface dust density corresponding to two different size distributions: $\mathcal{N}(s,t)\propto s^{-3}$ and  $\mathcal{N}(s,t)\propto s^{-4}$ and for scale heights of 1, 3 and 5 km. The limits of size range considered here are from 0.3 to 10 $\mu$m.
\end{table}

If we instead consider the populations with broader size distributions, our limits become much tighter. Table~\ref{tab:3} shows our surface
dust density limits obtained for the two different power-law size distributions (0.3 - 10 $\mu$m) and three different scale heights . There are two sets of values displayed in the table for the last three Almost Limb activities (1956A-1987A) with no Earthshine signal at all  and the last five activities (1918A-1987A) with negligible Earthshine present.
These limits on the surface dust density for particles larger than 0.3 $\mu$m are orders of magnitude above the values of 0.004 - 0.005 $m^{-3}$ measured by the LDEX experiment for the same size range \cite{horanyi2015permanent}. However, the UVS Almost-Limb observations probe a very different population from the LDEX measurements. The dust population seen by LDEX has a very large scale height, and the contribution of this diffuse cloud to any signal measured in the UVS Almost-Limb data is negligible. However, LDEX could only measure dust densities down to altitudes of around 3 km and so was insensitive to any additional dust population that might be confined to low altitudes, such as the extreme tails of the very low-altitude Lunar Horizon Glow seen by the Surveyor landers. Our upper limits constrain this additional population of low-altitude dust, and for a scale height of 1 km these limits correspond to densities of around 140 m$^{-3}$ near the surface, decreasing to values of around 0.006 $m^{-3}$ (comparable to the LDEX measurements) at altitudes around 10 km.

The dust constraints from the LADEE-UVS observations are therefore consistent with prior limits. Furthermore, since these measurements are sensitive to material within just a few kilometers of the surface, they can help constrain both impact-generated and electrostatically lofted dust in a region that had not been well constrained by prior measurements.

\section{Conclusions}
In summary, this analysis of the LADEE Almost Limb observations provide new constraints on the dust density at low altitudes above the Moon's surface. Specifically, the upper limit on the surface dust density for a mono-dispersed size distribution for a particle size of 0.1 $\mu m$ and a scale height of 1 km is 3.1 x 10$^4$ m$^{-3}$. The additional population of low-altitude dust for a scale height of 1 km and a size distribution proportional to $s^{-3}$ ranging from 0.3 to 10 $\mu$m (and $s^{-4}$) has an upper limit on dust density of around 140 m$^{-3}$ (and 190 m$^{-3}$) closer to the surface which reduces to around 6 x 10$^{-3}$ m$^{-3}$ (and 9 x 10$^{-3}$ m$^{-3}$) at altitudes reaching 10 km above the Moon's surface.

\appendix
\section{Formulas for computing predicted dust signals}
\label{appa}

Here we provide additional details about the formulas used to compute the predicted dust spectrum described in Section~\ref{sec:model} above. For this analysis, we use the optical theorem to approximate the scattering law for a dust grain  as:
\begin{equation}
\frac{dP}{d\Omega}=\frac{\pi F s^2}{4sin^2\theta} J_1^2(k\,s\,sin\theta) Q_{ext}^2(k\,s,m)
\end{equation}
where $F$ is the incident flux density, $s$ is the particle size (radius), $k$ is the wave number of the light, $\theta$ is the scattering angle, which is  12$^\circ$ for our observations, and the factor of  $Q_{ext}$ in Equation~\ref{eq:7} is defined as follows
\cite{van1957light}:
\begin{equation}
Q_{ext}=2-4e^{-\rho tan\beta}\frac{cos\beta}{\rho}\Bigg[sin(\rho-\beta)+\frac{cos\beta}{\rho} cos(\rho-2\beta)\Bigg]+4\frac{cos^2\beta}{\rho^2}
cos(2\beta),
\end{equation}
where $m$ is the complex refractive index of the particle ($m=m_r+im_i$), $\rho=2ks(m_r-1)$ and $tan\beta=m_i/(m_r-1)$. For these observations we assume a refractive index $m = 1.5 + i0.0005$ for a silicate-rich dust particle \cite{zubko2017reflectance}.

Since the dust above the Moon's surface is a collection of particles of different size ranges, we compute the unitless reflectance of this collection of particles with differential size distribution $\mathcal{N}(s,t)$ for wavelength channel $i$, with the following formula:
\begin{equation}
\Bigg[\frac{I}{F}(t)\Bigg]_{i,pred}=\frac{1}{F}\int_{s_{min}}^{s_{max}} \frac{dP_i}{d\Omega}\mathcal{N}(s,t)ds.
\label{eq:A3}
\end{equation}
For this analysis, we consider both mono-disperse size distributions and power-law size distributions with differential power-law indices of -3 and -4 and limits of 0.3 to 10 $\mu$m. 

For the specific case where,  $\mathcal{N}(s,t)=c_o(t) s^{-3}$, $c_o$ is a dimensionless function of time, but not particle size. The number of particles per unit area ($m^{-2}$), $N(t)$ is then given by the integral 
\begin{equation}
\label{eq:A4}
N(t)=c_o(t)\int_{s_{min}}^{s_{max}}s^{-3}ds.
\end{equation}
where $c_o(t)$ can be derived from equation~\ref{eq:A3} such that
\begin{equation}
c_o(t)=\frac{[\frac{I}{F}(t)]_{i,pred}}{{\frac{1}{F}\int_{s_{min}}^{s_{max}} \frac{dP_i}{d\Omega}s^{-3}ds}}.
\label{eq:A5}
\end{equation}
For reference, the value of the integral in the denominator for dust size from 0.3 ($s_{min}$) to 10 $\mu$m ($s_{max}$) for the wavelength range 250 - 300 nm is 21.0. This integral is worked out for all the wavelength ranges from 250 to 700 nm.

\begin{figure}[htbp]
\centering
\includegraphics[scale=0.5]{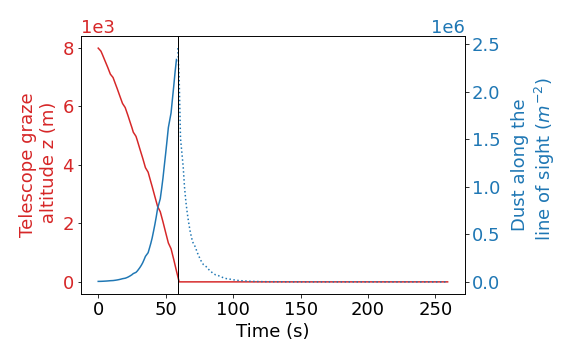}
\caption{The Telescope graze altitude ($z$) from Figure~\ref{fig:1}, as height of line of view at the point A (A in  Figure~\ref{fig:1}) on the surface is shown by the red line plotted against time for the Almost Limb activity 1969A. This altitude is given in the LBL file associated with each Almost Limb activity. It reduces to zero as the field of view touches the Moon's surface, which is indicated by the vertical line on the plot. The blue line shows the integrated dust density calculated using equation~\ref{eq:A6}. The amount of dust per unit area does increase as altitude decreases until the line of sight meets the surface.}
\label{fig:11}
\end{figure}

In order to compute the total amount of lunar dust visible at a given time along the line of sight, we assume an exponential dust profile defined by a nominal surface concentration $n_{o,nom}$ and scale height $H$. In this case, the total amount of dust along the line of sight is given by the following integral along the line of sight $x$:
\begin{equation}
N(t)=n_{o,nom}\int_{d}^{\infty}e^{-a(x,t)/H}dx
\label{eq:A6}
\end{equation}
where $N$ is the number of particles per unit area and $a(x,t)$ is the altitude above the lunar surface, and $d$ denotes the location where the light of sight crosses into sunlight (see Figure~\ref{fig:1}). Figure~\ref{fig:11} illustrates how this parameter ($N$) varies over the course of the observation, with a vertical line ($t_V$) denoting when the field of view hits the Moon's surface.

In order to determine the altitude $a(x,t)$ of the line of sight above the surface as a function of the coordinate $x$ along the line of sight at any given time t, we use the parameters defined in Figure~\ref{fig:1}.
The LBL data files associated with each spectrum gives the ``telescope graze altitude'' above the Moon's surface (denoted as $z$ in Figure~\ref{fig:1}) and the latitude and longitude of points $S$ (the spacecraft's location above the surface of the Moon) and $A$ (the place where the line of sight gets closest to the Lunar surface). Figure~\ref{fig:11} shows how the graze altitude $z$ changes over the course of the observation. It decreases gradually as the line of sight gets closer to the surface and reduced to zero as the line of sight hits the surface shown by the black vertical line on the plot.
The terminator position $T$ is calculated based on the sub-solar point on the Moon, which depends on the time at which the activity was performed.

Assuming the Moon to be a sphere, the height of the line of sight above the surface $a(x,t)$ can be written as the following function of $x$:
\begin{equation}
a(x,t)=\sqrt{x^2+(z+R)^2}-R
\label{eq:A7}
\end{equation}
where $R$ is the radius of Moon (1737.4 km).

The lower limit $d$ is calculated using the geometry shown in Figure~\ref{fig:1}. Considering the small triangle with legs $l$ and $d$, we see that
\begin{equation}
d=\frac{l}{\tan\epsilon_1}
\label{eq:d}
\end{equation} 
where $\epsilon_1$ is equal to the solar elevation angle of point $A$ (Figure~\ref{fig:1}):
\begin{equation}
\epsilon_1 = \sin^{-1}\Big[\sin\delta  \sin\phi + \cos\delta  \cos\phi  \cos(HRA) \Big]
\end{equation}
where $\delta$ is the sub-solar latitude angle, which depends on the time of the year and varies from (-0.5 to 0.5 deg for the Moon), $\phi$ is the latitude and HRA is the hour angle of point $A$ . Meanwhile, the length $l$ in Equation~\ref{eq:d} can be calculated from  $\epsilon_1$, $z$ and $R$ as
\begin{equation}
l=\frac{R}{\cos\epsilon_1}-(R+z).
\end{equation}
Note, the values of $l$, $\epsilon_1$ and $d$ are evaluated separated for each spectra based on data provided in each LBL file.

In Figure~\ref{fig:11}, we see how the integrated dust density increases with time as altitude decreases. This indicates that the amount of dust along the line of sight  is higher at lower altitude which is what one would expect. The slow decrease in $N$ after the line of view hits the surface (beyond the vertical line) is artificial because at that point the observable material would actually be blocked from view by the Moon.

Combining equation~\ref{eq:A4} and~\ref{eq:A6}, an expressions for predicted signal is obtained:
\begin{equation}
n_{o,nom}\int_{d}^{\infty}e^{-a(x,t)/H}dx=c_o(t)\int_{s_{min}}^{s_{max}} s^{-3}ds.
\end{equation}
Substituting in the above expression for $c_o(t)$ (equation~\ref{eq:A5}) and solving the integral for the size distribution $\mathcal{N}(s,t)\propto s^{-3}$, we find
\begin{equation}
n_{o,nom}\int_{d}^{\infty}e^{-(\sqrt{x^2+(z+R)^2}-R)/H}dx=\frac{\Big[\frac{I}{F}\Big]_{i,pred}}{\frac{1}{F}\int_{s_{min}}^{s_{max}} \frac{dP_i}{d\Omega}s^{-3}ds}\Bigg(\frac{s_{max}^2-s_{min}^2}{2s_{min}^2s_{max}^2}\Bigg).
\end{equation}
Solving this equation for $[I/F]_{i,pred}$ gives the predicted signal for a given particle size distribution ($\mathcal{N}(s,t)\propto s^{-3}$) and wavelength channel i:
\begin{equation}
\label{eq:A13}
\Big[\frac{I}{F}\Big]_{i,pred}=n_{o,nom}\Bigg[\frac{ 2s_{min}^2s_{max}^2}{s_{max}^2-s_{min}^2}\Bigg]\Bigg[\frac{1}{F}\int_{s_{min}}^{s_{max}} \frac{dP_i}{d\Omega}s^{-3}ds\Bigg]\int_{d}^{\infty}e^{-(\sqrt{x^2+(z+R)^2}-R)/H}dx.
\end{equation}
The above expression determines a predicted signal for a certain value of surface concentration ($n_{o,nom} = 10^3$ $m^{-3}$ and scale heights $H$ = 1, 3 and 5 km). 
 
Repeating these calculations for a size distribution proportional to $s^{-4}$, Equation~\ref{eq:A13} becomes 
\begin{equation}
\label{eq:A14}
\Big[\frac{I}{F}\Big]_{i,pred}=n_{o,nom}\Bigg[\frac{ 3s_{min}^3s_{max}^3}{s_{max}^3-s_{min}^3}\Bigg]\Bigg[\frac{1}{F}\int_{s_{min}}^{s_{max}} \frac{dP_i}{d\Omega}s^{-4}ds\Bigg]\int_{d}^{\infty}e^{-(\sqrt{x^2+(z+R)^2}-R)/H}dx
\end{equation}
and for a mono-disperse size distribution (0.07 to 1 $\mu$m)
Equation~\ref{eq:A13} changes to
\begin{equation}
\label{eq:A15}
\Big[\frac{I}{F}\Big]_{i,pred}=\frac{n_{o,nom}\pi s^{2}}{4sin^2\theta}J_1^2(k\,s\,sin\theta) Q_{ext}^2(k\,s,m)\int_{d}^{\infty}e^{-(\sqrt{x^2+(z+R)^2}-R)/H}dx.
\end{equation}

\newpage
\section{Plots showing the measured signals for all observations after spectral filtering}
\label{appn2}

\begin{figure}[htbp]
\centering
\includegraphics[scale=0.56]{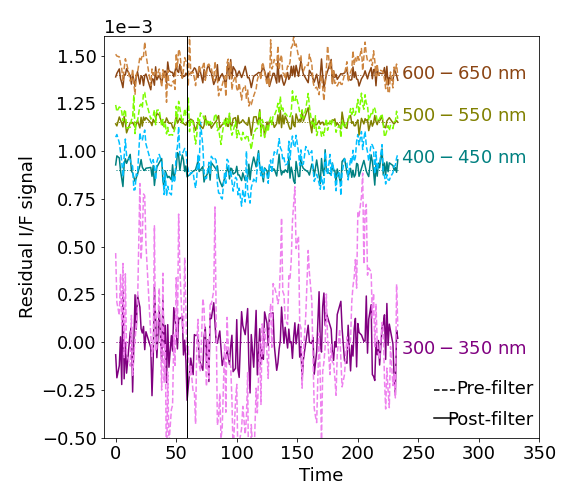}
\caption{Corrected signal (in solid) for the set "1918A", after spectral filtering has been applied to each wavelength range and the original spectrum (in dotted), plotted against time. An offset has been added here too to view each wavelength range distinctly.}
\label{fig:12}
\end{figure}

\begin{figure}[bp]
\centering
\includegraphics[scale=0.7]{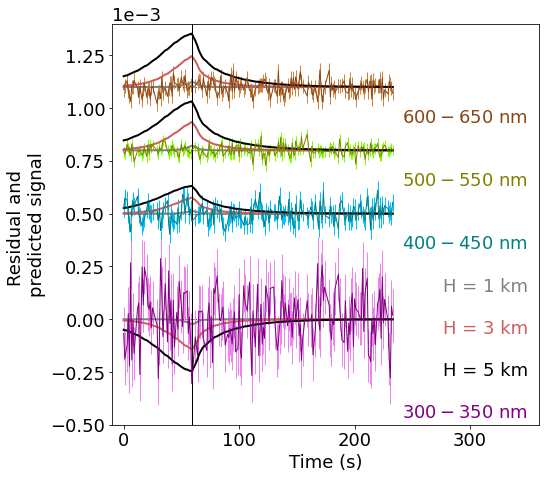}
\caption{This figure shows the filtered residual signal for "1918A" for each wavelength range. The signal predicted using the Fraunhofer model and the exponential density profile is plotted over the residual signal for a scale heights of 1, 3 and 5 km in grey, lightcoral and black respectively. An offset has been added to view the signals distinctly.}
\label{fig:13}
\end{figure}

\begin{figure}[htbp]
\centering
\includegraphics[scale=0.7]{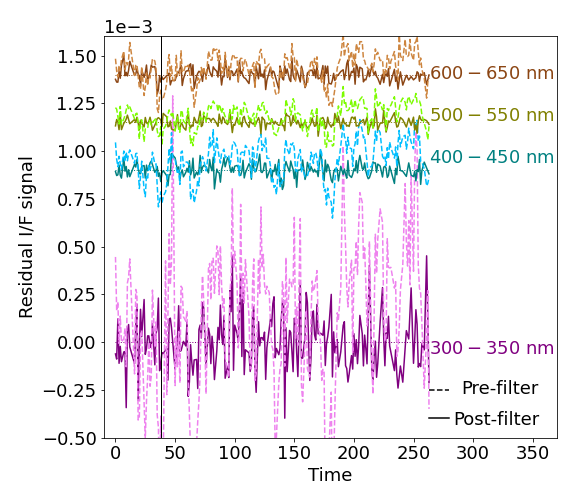}
\caption{Corrected signal (in solid) for the set "1929A", after spectral filtering has been applied to each wavelength range and the original spectrum(in dotted), plotted against time. An offset has been added here too to view each wavelength range distinctly.}
\label{fig:14}
\end{figure}

\begin{figure}[htbp]
\centering
\includegraphics[scale=0.7]{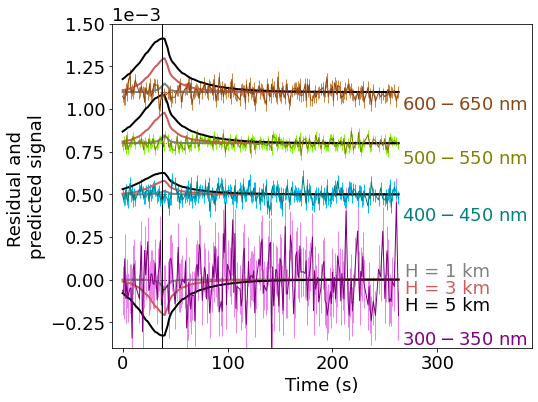}
\caption{This figure shows the filtered residual signal for "1929A" for each wavelength range. The signal predicted using the Fraunhofer model and the exponential density profile is plotted over the residual signal for a scale heights of 1, 3 and 5 km in grey, lightcoral and black respectively. An offset has been added to view the signals distinctly.}
\label{fig:15}
\end{figure}

\begin{figure}[htbp]
\centering
\includegraphics[scale=0.7]{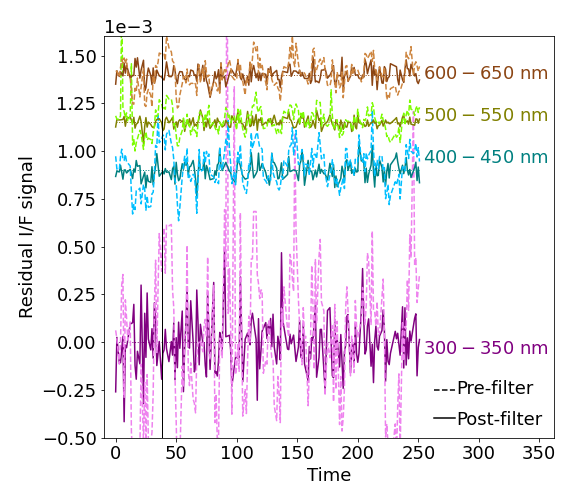}
\caption{Corrected signal (in solid) for the set "1956A", after spectral filtering has been applied to each wavelength range and the original spectrum(in dotted), plotted against time. An offset has been added here too to view each wavelength range distinctly.}
\label{fig:16}
\end{figure}

\begin{figure}[htbp]
\centering
\includegraphics[scale=0.7]{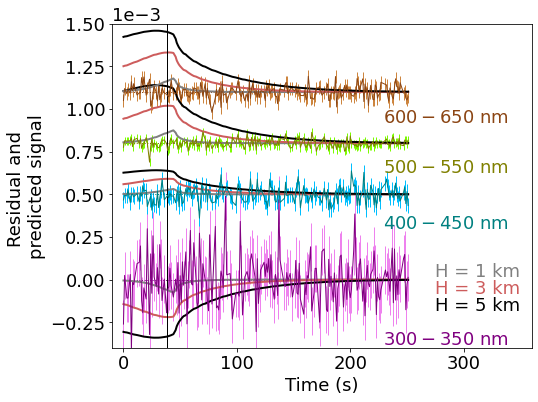}
\caption{This figure shows the filtered residual signal for "1956A" for each wavelength range. The signal predicted using the Fraunhofer model and the exponential density profile is plotted over the residual signal for a scale heights of 1, 3 and 5 km in grey, lightcoral respectively. An offset has been added to view the signals distinctly.}
\label{fig:17}
\end{figure}

\begin{figure}[htbp]
\centering
\includegraphics[scale=0.7]{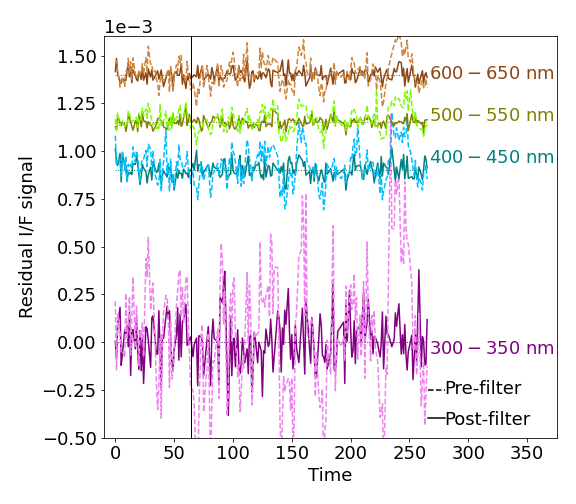}
\caption{Corrected signal (in solid) for the set "1987A", after spectral filtering has been applied to each wavelength range and the original spectrum(in dotted), plotted against time. An offset has been added here too to view each wavelength range distinctly.}
\label{fig:18}
\end{figure}

\begin{figure}[htbp]
\centering
\includegraphics[,scale=0.7]{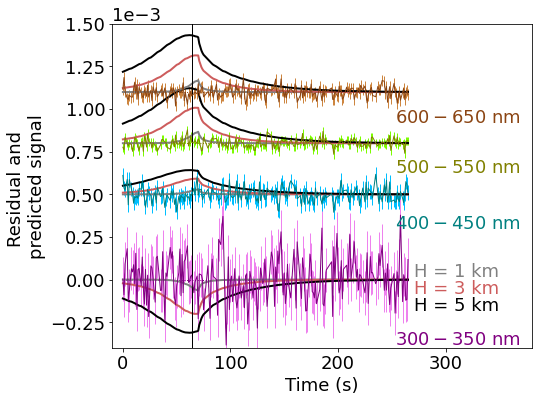}
\caption{This figure shows the filtered residual signal for "1987A" for each wavelength range. The signal predicted using the Fraunhofer model and the exponential density profile is plotted over the residual signal for a scale heights of 1, 3 and 5 km in grey,lightcoral and black respectively. An offset has been added to view the signals distinctly.}
\label{fig:19}
\end{figure}

\clearpage

\section{Tables of fit parameters for all observations}
\label{appn3}

\begin{figure}[t]
\centering
\includegraphics[scale=0.499]{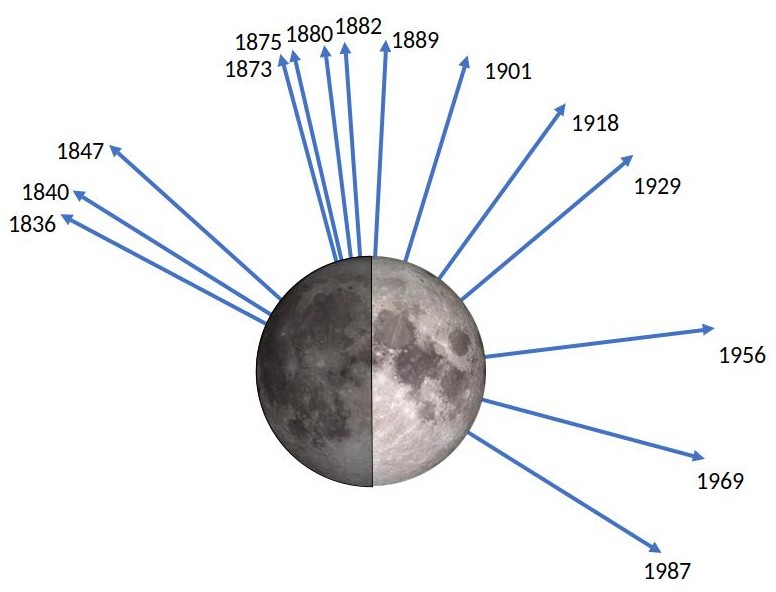}
\caption{This figure shows the location of Earth (arrows point to Earth) during the month of April 2014 while the Almost Limb activities were recorded by LADEE-UVS. It shows how the first observations are affected due to Earthshine of the dark side of the Moon while the last 5 observations have negligible amount of Earthshine.}
\label{fig:20}
\end{figure}

Table~\ref{tab:4} shows the slope between the residual and predicted signal and the standard deviation of this slope is calculated for each wavelength range. A weighted average of these slopes using their respective standard deviation as weights is computed for each observation. The weighted averaged slope with error bars at scale height of 1, 3 and 5 km are shown in this table. The intercept for this case is reduced to zero. Table~\ref{tab:4} and Figure~\ref{fig:20} also provide the Earth-Moon position details for these observations. The observed locations are farther and farther away from Earthshine horizon for the later data sets. 
\begin{table}[tbh]
\caption{Observation geometry and fitted slopes for all Almost Limb activities}
\centering
\begin{tabular} {c|c|c|c|c|c|c}
\hline\hline
\multirow{4}{4em}{Activity name} & \multirow{4}{*}{H = 1 km} & \multirow{4}{*}{H = 3 km} & \multirow{4}{*}{H = 5 km} & \multirow{4}{5em}{Days since Full Earth / New Moon} & \multirow{3}{6em}{Difference between Sub-Earth and observed point} & \multirow{3}{3em}{Earth-Moon Phase angle} \\[12ex]
\cline{2-4}
 & Slope & Slope & Slope & & & \\[0.5ex]
 \hline
1836A & -377.33 $\pm$ 61.20 & -4.01 $\pm$ 0.28 & -1.25 $\pm$ 0.05 & 1.72 & 52.89& 20.6\\
1840A & -585.04 $\pm$ 60.39 & -4.83 $\pm$ 0.33 & -1.28 $\pm$ 0.08 & 2.72 & 48.05 & 32.6\\
1847A & -339.02 $\pm$ 29.87 & -3.27 $\pm$ 0.22 & -0.92 $\pm$ 0.05 & 3.72 & 38.59 & 44.6\\
1873A & -85.74 $\pm$ 13.06 & -0.72 $\pm$ 0.09 & -0.20 $\pm$ 0.02 & 5.72 & 6.20 & 68.6\\
1875A & -25.95 $\pm$ 10.67 & -0.30 $\pm$ 0.11 & -0.08 $\pm$ 0.36 & 6.72 & 3.55 & 80.6\\
1880A & -21.75 $\pm$ 12.05 & -0.48 $\pm$ 0.11 & -0.15 $\pm$ 0.03 & 6.72 & -2.61 & 80.6\\
1882A & -84.67 $\pm$ 21.45 & -0.59 $\pm$ 0.12 & -0.14 $\pm$ 0.03 & 6.72 & -4.36 & 80.6\\
1889A & 12.72 $\pm$ 23.70 & 0.19 $\pm$ 0.13 & 0.05 $\pm$ 0.03 & 7.72 & -12.25 & 92.6\\
1901A & -64.00 $\pm$ 32.74 & -0.60 $\pm$ 0.18 & -0.16 $\pm$ 0.04 & 8.72 & -26.17 & 104.6\\
1918A & 0.08 $\pm$ 0.34 & 0.03 $\pm$ 0.03 & 0.01 $\pm$ 0.01 & 10.72 & -40.18 & 128.6\\
1929A & -0.09 $\pm$ 0.24 & 0.02 $\pm$ 0.02 & 0.01 $\pm$ 0.01 & 11.72 & -53.23 & 140.6\\
1956A & 0.02 $\pm$ 0.09 & -0.01 $\pm$ 0.03 & 0.00 $\pm$ 0.02 & 14.72 & -85.51 & 176.6\\
1969A & 0.81 $\pm$ 0.12 & 0.00 $\pm$ 0.01 & -0.01 $\pm$ 0.01 & 16.72 & -107.27 & 200.6\\
1987A & 0.20 $\pm$ 0.14 & 0.02 $\pm$ 0.02 & 0.01 $\pm$ 0.01 & 17.72 & -124.76 & 212.6\\
[1ex]
\hline
\label{tab:4}
\end{tabular}
\end{table}

\newpage

\newpage
Table~\ref{tab:5} shows the slope  and intercept between the residual, $D_{i,est}(t)$ and predicted signal, $P_{i,est}(t)$ and the standard deviation of this slope and intercept are calculated for each wavelength range $i$. Weighted average of these slopes and intercepts using their respective standard deviation as weights is computed for each observation. The weighted averaged slope and intercepts with error bars at scale height of 1, 3 and 5 km are shown in this table. The intercept for this case is not reduced to zero. And the Earth-Moon position details during the course of these observations. The observed locations are farther and farther away from Earthshine horizon for the later data sets.
\begin{sidewaystable}[t]
\small
\caption{Slopes and Intercepts for all Almost Limb activities}
\label{tab:5}
\centering
\begin{adjustbox}{scale=0.8,center}
\begin{tabular}{c|c|c|c|c|c|c|c|c|c}
\hline\hline
\multirow{2}{*}{Activity name}& \multicolumn{2}{c|}{H = 1 km} & \multicolumn{2}{c|}{H = 3 km}& \multicolumn{2}{c|}{H = 5 km} & H = 1 km & H = 3 km & H = 5 km  \\
\cline{2-10}
 & Slope  & Intercept x$10^{-5}$ & Slope & Intercept x$10^{-5}$ & Slope & Intercept x$10^{-5}$ & Slope & Slope & Slope\\
 \hline
1836A & 54.50 $\pm$ 35.94 & -4.64 $\pm$ 0.13 & 0.59 $\pm$ 0.23 & -4.83 $\pm$ 0.17 & 0.22 $\pm$ 0.08 & -5.11 $\pm$ 0.25 & -377.33 $\pm$ 61.20 & -4.01 $\pm$ 0.28 & -1.25 $\pm$ 0.05\\
1840A & 225.74 $\pm$ 52.45 & -4.41 $\pm$ 0.13 & 1.13 $\pm$ 0.29 & -4.60 $\pm$ 0.17 & 0.34 $\pm$ 0.09 & -4.88 $\pm$ 0.24 & -585.04 $\pm$ 60.39 & -4.83 $\pm$ 0.33 & -1.28 $\pm$ 0.08 \\
1847A & 22.19 $\pm$ 27.64 & -3.49 $\pm$ 0.13 & 0.46 $\pm$ 0.22 & -3.72 $\pm$ 0.18 & 0.18 $\pm$ 0.08 & -4.01 $\pm$ 0.29 & -339.02 $\pm$ 29.87 & -3.27 $\pm$ 0.22 & -0.92 $\pm$ 0.05 \\
1873A & -26.45 $\pm$ 20.21 & -0.71 $\pm$ 0.25 & -0.55 $\pm$ 0.45 & 0.02 $\pm$ 0.75 & -0.30 $\pm$ 0.25 & 1.04 $\pm$ 1.50 & -85.74 $\pm$ 13.06 & -0.72 $\pm$ 0.09 & -0.20 $\pm$ 0.02\\
1875A & 20.93 $\pm$ 14.69 & -0.78 $\pm$ 0.21 & 0.48 $\pm$ 0.36 & -1.34 $\pm$ 0.59 & 0.27 $\pm$ 0.20 & -2.12 $\pm$ 1.17 & -25.95 $\pm$ 10.67 & -0.30 $\pm$ 0.11 & -0.08 $\pm$ 0.36 \\
1880A & 32.60 $\pm$ 14.82 & -0.85 $\pm$ 0.18 & 0.61 $\pm$ 0.28 & -1.44 $\pm$ 0.42 & 0.32 $\pm$ 0.15 & -2.24 $\pm$ 0.81 & -21.75 $\pm$ 12.05 & -0.48 $\pm$ 0.11 & -0.15 $\pm$ 0.03 \\
1882A & -22.79 $\pm$ 27.10 & -0.70 $\pm$ 0.20 & -0.25 $\pm$ 0.35 & -0.45 $\pm$ 0.46 & -0.12 $\pm$ 0.17 & -0.10 $\pm$ 0.89 & -84.67 $\pm$ 21.45 & -0.59 $\pm$ 0.12 & -0.14 $\pm$ 0.03\\
1889A & -6.46 $\pm$ 26.50 & 0.23 $\pm$ 0.14 & 0.02 $\pm$ 0.25 & 0.24 $\pm$ 0.23 & 0.02 $\pm$ 0.10 & 0.24 $\pm$ 0.40 & 12.72 $\pm$ 23.70 & 0.19 $\pm$ 0.13 & 0.05 $\pm$ 0.03 \\
1901A & -30.75 $\pm$ 34.97 & -0.34 $\pm$ 0.13 & -0.39 $\pm$ 0.23 & -0.22 $\pm$ 0.15 & -0.14 $\pm$ 0.07 & -0.06 $\pm$ 0.22 & -64.00 $\pm$ 32.74 & -0.60 $\pm$ 0.18 & -0.16 $\pm$ 0.04\\
1918A & -0.20 $\pm$ 0.40 & 0.18 $\pm$ 0.22 & 0.02 $\pm$ 0.05 & 0.02 $\pm$ 0.33 & 0.02 $\pm$ 0.04 & -0.12 $\pm$ 0.53 & 0.08 $\pm$ 0.34 & 0.03 $\pm$ 0.03 & 0.01 $\pm$ 0.01 \\
1929A & -0.42 $\pm$ 0.29 & 0.41 $\pm$ 0.24 & -0.06 $\pm$ 0.05 & 0.59 $\pm$ 0.39 & -0.05 $\pm$ 0.04 & 0.85 $\pm$ 0.7 & -0.09 $\pm$ 0.24 & 0.02 $\pm$ 0.02 & 0.01 $\pm$ 0.01\\
1956A & 0.07 $\pm$ 0.19 & -0.26 $\pm$ 0.55 & -0.07 $\pm$ 0.15 & 0.88 $\pm$ 2.70 & -0.60 $\pm$ 0.39 & 16.1 $\pm$ 12.3 & 0.02 $\pm$ 0.09 & -0.01 $\pm$ 0.03 & 0.00 $\pm$ 0.02\\
1969A & 0.16 $\pm$ 0.16 & -0.19 $\pm$ 0.24 & 0.03 $\pm$ 0.04 & -0.37 $\pm$ 0.58 & 0.02 4$\pm$ 0.05 & -0.56 $\pm$ 1.33 & 0.81 $\pm$ 0.12 & 0.00 $\pm$ 0.01 & -0.01 $\pm$ 0.01 \\
1987A & 0.13 $\pm$ 0.17 & 0.09 $\pm$ 0.20 & 0.01 $\pm$ 0.03 & -0.14$\pm$ 0.34 & 0.00 $\pm$ 0.03 & 0.23 $\pm$ 0.63 & 0.20 $\pm$ 0.14 & 0.02 $\pm$ 0.02 & 0.01 $\pm$ 0.01\\
[1ex]
\hline 
\end{tabular}
\end{adjustbox}
\end{sidewaystable}

\clearpage
\newpage
\acknowledgments
This work was supported by a Lunar Data Analysis Program Grant NNX15AV54G. 

The primary data used in this analysis is all available on the Planetary Data System and is cataloged under \citeA{https://doi.org/10.17189/1518945}.

Datasets for this research are available in this in-text data
citation references: \cite{https://doi.org/10.7923/xnej-7h18}, [with license: \url{https://creativecommons.org/licenses/by-nc-sa/4.0/}, and the access restrictions: Creative Commons Attribution Non Commercial Share Alike 4.0 International] at the University of Idaho' Institutional repository.

The type of data files used in this analysis include .sav and .txt files. These files are read using .C (C code) and .ipynb (Python notebooks) code files.
The processed data and the model data along with the codes and tools have been uploaded to the following \citeA{https://doi.org/10.7923/xnej-7h18}

\bibliography{Bibliography}
\end{document}